\documentclass[9pt,twocolumn,twoside]{pnas-new}

\templatetype{pnasresearcharticle} 

\title{Scale-free Resilience of Real Traffic Jams}

\author[a,b]{Limiao Zhang}
\author[a,b]{Guanwen Zeng} 
\author[a,b,1]{Daqing Li}
\author[c,1]{Hai-Jun Huang}
\author[d,e,1]{H. Eugene Stanley}
\author[f]{Shlomo Havlin}

\affil[a]{School of Reliability and Systems Engineering, Beihang University, Beijing 100191, China}
\affil[b]{Science and Technology on Reliability and Environmental Engineering Laboratory, Beijing 100191, China}
\affil[c]{School of Economics and Management, Beihang University, Beijing 100191, China}
\affil[d]{Center for Polymer Studies, Boston University, Boston, MA 02215}
\affil[e]{Department of Physics, Boston University, Boston, MA 02215}
\affil[f]{Department of Physics, Bar-Ilan University, Ramat-Gan 52900, Israel}

\correspondingauthor{\textsuperscript{1}To whom correspondence should be addressed. \\*E-mail: daqingl@buaa.edu.cn, haijunhuang@buaa.edu.cn, hes@bu.edu}


\begin{abstract}
The concept of resilience can be realized in natural and engineering systems, representing the ability of system to adapt and recover from various disturbances. Although resilience is a critical property needed for understanding and managing the risks and collapses of transportation system, an accepted and useful definition of resilience for urban traffic as well as its statistical property under perturbations is still missing. Here we define city traffic resilience based on the spatio-temporal clusters of congestion in real traffic, and find that the resilience follows a scale free distribution in two-dimensional city road networks and one-dimensional highways, with different exponents, but similar exponents in different days and different cities. The traffic resilience is also revealed to have a novel scaling relation between the cluster size of the spatio-temporal jam and its recovery duration, independent of microscopic details. Our findings of universal traffic resilience can provide indication towards better understanding and designing these complex engineering systems under internal and external disturbances.
\end{abstract}

\begin{document}

\verticaladjustment{-2pt}

\maketitle
\thispagestyle{firststyle}
\ifthenelse{\boolean{shortarticle}}{\ifthenelse{\boolean{singlecolumn}}{\abscontentformatted}{\abscontent}}{}

\dropcap{I}ncreasing traffic congestion is an inescapable problem due to enhanced urbanization and growing metropolitan cities all over the world, from Los Angeles to Tokyo and from Cairo to Beijing \cite{r1}, leading to potential high economic and social losses. Under various internal or external perturbations, ranging from a local flow fluctuation to a broken-down traffic light and up to extreme weather conditions, a small jam can develop into a large-scale congestion in a domino-like cascading process \cite{r2}. Given the uncertainty of disruptive system failures, the concept of resilience describes the system ability to withstand possible perturbations and recover to an acceptable functional level. Since Holling’s definition in ecology \cite{r3}, the resilience framework has been developed and applied in many disciplines, ranging from climate, economics to social science \cite{r4,r5,r6,r7,r8,r9,r10,r11,r12}. System resilience across different domains usually depends on its absorptive capacity, adaptive capacity and restorative capacity \cite{r13}. Accordingly, system adaptation and recovery process in various critical infrastructures including transportation have attracted much attention recently \cite{r14,r15,r16,r17}. Especially, a resilient transportation system in the future smart city era could improve significantly life quality, the development of economic society and reducing environmental pollutions \cite{r18}.

Transportation systems with network topology, as one of the critical infrastructures, serve as the lifeline for national economics and stability. System resilience has been studied in different traffic systems including city roads, metro system, freight transportation and aviation network \cite{r19,r20,r21,r22,r23,r24,r25}. While different methods have been proposed to evaluate and improve the resilience of transportation and other infrastructures, the resilience metric is mainly based on a dimensionless indicator. Chang and Shinozuka \cite{r26} introduced this resilience measurement that relates expected losses in future disasters to a community seismic performance objectives. It has been proposed to define for earthquakes the measurement of resilience as the change in system performance over time \cite{r27}, which is the well-known resilience triangle. It measures the resilience loss of a community due to an earthquake using,
\begin{align*}
RL=\int_{t_{0}}^{t_{1}}[{\color{black}1}-Q(t)]dt.\numberthis \label{eqn:powerlaw}
\end{align*}
Here Q(t) represents the service quality (ranged between 0\% to 100\%) of the community, which starts to decrease at $t_0$ and may {\color{black}{return}} to its normal state (100\%) at $t_1$. Though this method is presented in the context of earthquakes, the concept has been widely applied for other scenario-specific system performance under various disturbances \cite{r28,r29}. Meanwhile, although critical for understanding and improving the system robustness and vulnerability \cite{r30,r31,r32,r33,r34,r35,r36,r37} , the network topology has rarely been considered in resilience studies of critical infrastructures and other complex systems. Since traffic system in a city has a typical network structure and its resilience evolves both, in space and time, the above-mentioned dimensionless resilience indicators and relevant studies may have missed the spatio-temporal properties of {\color{black}{system adaptation and recover}} in these critical infrastructure networks.

Composed of a very large number of strongly interacting subunits, transportation systems are usually running out of equilibrium states with unpredictable outcome of cascading failures \cite{r38}. Due to the longstanding debate whether system resilience is intrinsic \cite{r39}, it is critical yet unknown if such systems with numerous interacting subunits have universal resilient behavior that is independent of microscopic details. Here we propose a new spatio-temporal resilience measure, incorporating both spatial and temporal features of system adaptation and recover, in order to explore the possible universality features of traffic resilience. With extensive real traffic data, we find novel scaling laws from scale free distributions for the traffic resilience and recovery duration. Our definition and results demonstrate and support the existence of intrinsic behavior behind traffic resilience independent of microscopic details. These scaling laws hold for different size scales of traffic jammed clusters, which can help to predict system restoration behaviors and develop corresponding resilience management methods.

\section*{Results}

\begin{figure}
\centering
\includegraphics[width=1.0\linewidth]{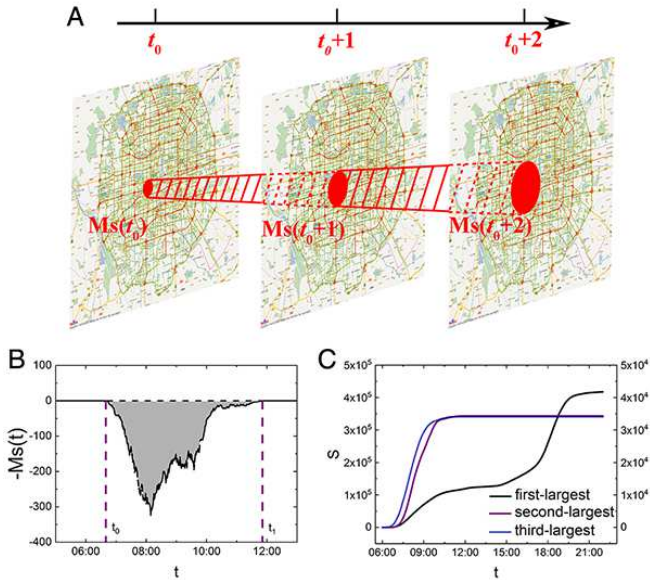}
\caption{\textbf{Traffic resilience defined based on spatio-temporal jammed clusters.}  (\emph{A}) Illustration of the evolution of a jammed cluster in a city. Red links are considered congested. All red links in the shadow belong to the same jammed cluster. (\emph{B}) The cross section area \emph{$M_s{(t)}$} of the second-largest jammed cluster on Oct. 26th 2015 in Beijing. Since the resilience is reduced during the jam, we plot the negative of \emph{$M_s{(t)}$} as a function of time, and traffic resilience can be represented by the grey area. The grey area is the size of the spatio-temporal jammed cluster (\emph{S}), shown in red in  (\emph{A}). Time span between \emph{$t_0$} and \emph{$t_1$}  represents its recovery time (\emph{T}=\emph{$t_1$}-\emph{$t_0$}+1). (\emph{C}) The cluster sizes of the first, second and third-largest jammed cluster on Oct. 26th 2015 in Beijing as a function of time (the second and third largest clusters sizes are given on the right axis scale).}\label{fig:one}
\end{figure}

Our study uses real traffic GPS data of Beijing and Shenzhen, which are two of megacities that suffer from most severe traffic jams world-wide and particularly in China. Complex road topology, large traffic flow and various perturbations as well as the availability of big data make these two megacities ideal for urban traffic research of resilience. The static road network in Beijing contains over 39,000 road segments (links) and 27,000 intersections (nodes), while Shenzhen traffic network contains about 18,000 road segments (links) and 12,000 intersections (nodes). The data set covers GPS velocity records in both cities for 30 days during October 2015 with resolution of one minute. A dynamical traffic network can be constructed based on road topology information and high resolution of evolving traffic velocity data. Each road in the network has a velocity $v_i$  (km/h) and a given velocity threshold $p_i$ is determined to judge the traffic availability of this road (detailed thresholds for different roads are shown in {\color{black}{SI Appendix,}} Table S1). {\color{black}{We also tested the influence of the threshold, and find that our results are insensitive to the thresholds (for details in SI Appendix, Fig. S1-S3).}} Then, roads with real-time velocity $v_i$ below the threshold are regarded as congested. Specifically, the links in the jammed cluster at a given time represent congestion roads, while nodes in the jammed cluster are the intersections between these congested roads. Considering together the temporal evolution, as well as the two-dimensional spatial traffic network, we can regard the jam as a three-dimensional spatio-temporal network cluster. Accordingly, a three-dimensional (two of space and one of time) cluster can be constructed representing the same jam during its entire lifetime. The three-dimensional jammed cluster is demonstrated in Fig. \ref{fig:one}\emph{A}, where all red links in the shadow belong to the same jammed cluster. Note that the connected clusters here do not necessary mean that any roads within a connected cluster are spatially connected at a given time instant. When a jammed cluster splits into two or more sub-clusters at a certain instant, all links and nodes in the sub-clusters still belong to the same three-dimensional cluster due to their temporal connection. Our definition of jammed clusters intuitively reflects the spatio-temporal propagation and dissolution of traffic jams, instead of earlier dimensionless resilience indicators.

We define the resilience based on the three-dimensional cluster size, using conceptually Eq. (2) as follows. For each jammed cluster during the observed period, e.g. from 6:00 to 22:00, the number of its links (roads) at a snapshot of the temporal layer \emph{t}, \emph{$M_s{(t)}$} varies with time. Thus \emph{$M_s{(t)}$} can be regarded as the cross section area of the jammed cluster at time \emph{t}. Larger \emph{$M_s{(t)}$} means that more roads are congested at snapshot \emph{t}.  {\color{black}{The maximal cross section of the spatio-temporal congestion cluster in Beijing is also plotted (see SI Appendix, Fig. S4).}} We evaluate the resilience performance of the traffic network by analyzing the evolution and statistics of \emph{$M_s{(t)}$}. For example, the time evolution of \emph{$M_s{(t)}$} of the second-largest jammed cluster on Oct. 26th 2015 is demonstrated in Fig. \ref{fig:one}\emph{B}. The time span between \emph{$t_0$} and \emph{$t_1$}, which is the lifetime of this jammed cluster, is defined as the recovery duration (\emph{T}=\emph{$t_1$}-\emph{$t_0$}+1). The recovery duration reflects how long it takes for this jammed cluster to recover from the beginning of congestion. We define the cluster size, \emph{S}, as the total number of links (roads) in the jammed cluster during its recovery time as,
\begin{align*}
S=\int_{t_{0}}^{t_{1}}M_s{(t)}dt.\numberthis \label{eqn:powerlaw}
\end{align*}
The cluster sizes of the first three largest jammed clusters on Oct. 26th 2015 in Beijing as a function of time are demonstrated in Fig. \ref{fig:one}\emph{C}. The cluster size naturally represents the loss of resilience in the traffic network. Eq. (2) does not only characterize the propagation of congestion in spatial dimension, but also includes the duration of congestion. Thus, the larger the jammed cluster size is, the less resilient the traffic system should be regarded. The shadow area shown in Fig. \ref{fig:one}\emph{B} represents therefore this loss of traffic resilience. To show the daily variations in the cluster sizes, we plot the size of the first three largest clusters as a function of date in Beijing (see {\color{black}{SI Appendix, Fig. S5}}). The largest cluster sizes are found obviously smaller in holidays (Oct. 1st - Oct. 7th {\color{black}{2015}}), due to the less traffic demand compared to normal workdays. {\color{black}{Note that when two (or more) jammed clusters merge, they will be regarded as a single three-dimensional cluster. We update the information of all links and nodes in the sub-clusters and identify them as a single jammed cluster.}}

\begin{figure}
\centering
\includegraphics[width=1.0\linewidth]{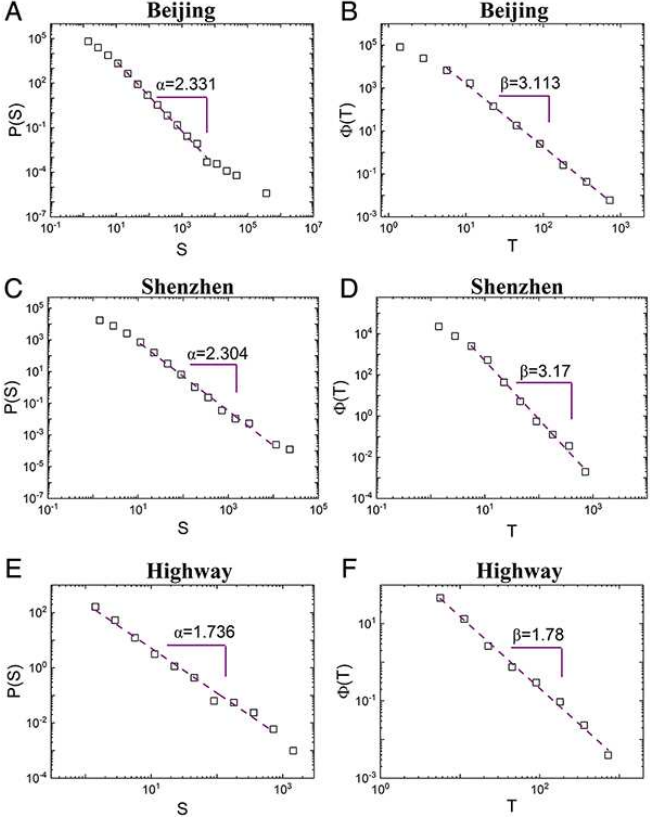}
\caption{\textbf{Scale free distributions of traffic resilience.}  (\emph{A}) The distribution of jammed cluster size. (\emph{B}) The distribution of recovery duration. (\emph{A}) and (\emph{B}) are typical results based on city traffic data in Beijing on Oct. 26th 2015. (\emph{C}) and (\emph{D}) are typical results based on city traffic data in Shenzhen on Oct. 26th 2015. (\emph{E}) and (\emph{F}) are typical results based on traffic data of Beijing-Shenyang Highway on Oct. 1st 2015. The results are analyzed by logarithmic bins, and plotted in double-logarithmic axis.}\label{fig:two}
\end{figure}

Next, we explore the distributions of cluster sizes and recovery duration in a typical day. The results on Monday Oct. 26th 2015 in Beijing and Shenzhen are shown in Fig. \ref{fig:two}. The distribution of cluster sizes shows a scale-free property, i.e., a power law scaling,
\begin{align*}
P(S) \sim S^{-\alpha },\numberthis \label{eqn:powerlaw}
\end{align*}
with an exponent $\alpha$ close to 2.3 in both cities. The power law distribution of cluster size suggests that although most of the congestions are of small-scales, there exist every day congestions of sizes at all of the scales including extremely large spatio-temporal scale. With the signal-control based traffic management, small jams due to fluctuating traffic demand or accidents in a city will usually shrink and dissolve after a short timespan. However, if the traffic supply under real-time management cannot meet the increasing traffic demand, traffic jam will grow to a large scale and take more time to recover. These two behaviors compete in different scales in the city and possibly lead to the scale-free distribution of traffic resilience. This also suggests that under different level of internal or external perturbation, transportation system has the same response distribution described by one scaling function.

\begin{figure} 
\centering
\includegraphics[width=1.0\linewidth]{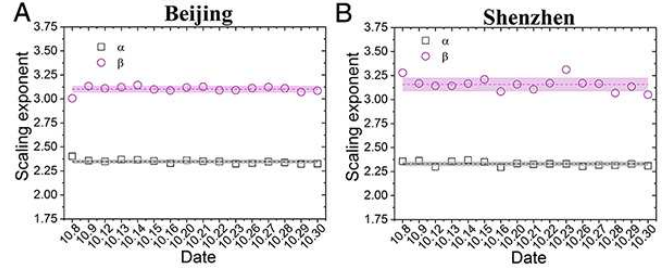}
\caption{\textbf{Scaling exponents of the scale-free distributions of cluster size and recovery durations as a function of date}  (\emph{A}) in Beijing,  (\emph{B})  in Shenzhen.}\label{fig:three}
\end{figure}

We also find that the cluster size distribution in both cities follows a very similar power law ($\alpha$ = 2.34 ± 0.02) for all observed workdays (see Fig. \ref{fig:three}). The high-quality scaling laws found here in different cities and different periods highly {\color{black}{suggest}} that resilience defined here may reflect an intrinsic property of urban traffic, independent on the microscopic traffic details that change from day to day and from city to city. Since all sizes seem to follow the same scaling law, a unified resilience management may exist for different sizes and locations of jams.

Next, we analyze and explore the scaling properties of the recovery duration in traffic congestion. In Fig. \ref{fig:two}\emph{B}, we show the distribution in a typical day in Beijing. It is found that the recovery duration of jammed clusters follows a distinct power-law distribution,
\begin{align*}
\Phi (T) \sim T^{-\beta},\numberthis \label{eqn:powerlaw}
\end{align*}
with an exponent $\beta$. Furthermore, similar results for the scaling exponent $\beta$, are also found for another city of Shenzhen (Fig. \ref{fig:two}\emph{D}). In these two megacities, the power law distributions for system recover are similar in all of observed days with $\beta$ = 3.13 ± 0.06 (see Fig. \ref{fig:three}). Under different possible perturbations, there seems to exist all scales of recovery duration including some cases of very long recovery duration, but all (short, medium and long recovery durations) follow the same scaling law. This scaling law {\color{black}{enables}} us to understand the common recovery mechanism for different sizes of jammed clusters, which would be helpful for mitigation guidance. 

Surprisingly, the power law exponents of resilience cluster size and recovery duration distributions are found stable in different days of two cities during the observed period (see Fig. \ref{fig:three}). The appearance of the power law and its stability in different working days for a city is probably due to the self-organized nature \cite{r40} of traffic flow and corresponding optimized management in urban traffic. On one hand, a large number of vehicles rush into the road network during peak hours, which fluctuates from day to day. Once the traffic flow returns to normal status, congestions disappear spontaneously. On the other hand, corresponding traffic control strategies such as traffic diversion, traffic lights and speed limitation are applied to alleviate specific traffic jams on a given day and pursue the system efficiency \cite{r41, r42}. All of these push the system towards its intrinsic operational limits, which might contribute to our findings of robust scale-free distributions of cluster sizes and recovery duration.  {\color{black}{Since our transportation system of a city is a large system with relatively similar daily flow demand and corresponding traffic control strategy, this is similar to the sand-pile model for self-organized criticality \cite{r40}, where the critical state is also robust under perturbation. Our definition of spatio-temporal resilience in some sense measures the spatio-temporal scale range of the attraction basin.}}

{\color{black}{Our traffic system can be seen as analogous to the sand-pile model since ‘particles’ (cars) are continuously added into the transportation system in a city starting early every morning. Then a local perturbation of jam may develop and spread to neighboring sites like a domino effect, forming congestion of all sizes as a result of the self-organized criticality similar to the sand pile model. This self-organized behavior generates spatial self-similar structures and temporal correlations across a broad range of scales, similar to the sand pile model. Here we found that the spatial and temporal scaling behaviors interact and form a scale-free size distribution in the \emph{d}+1 resilience clusters.}} The universal features suggested by the scale-free nature of traffic resilience should as usually found to depend on a few macroscopic variables including network dimension \cite{r43} and total traffic demand. To test this hypothesis, we also analyzed the traffic data of Beijing-Shenyang Highway between Oct. 1st 2015 and Oct. 7th 2015. This observed timespan is the National Day holiday in China, during which the highway is usually under heavy traffic pressure. A highway can be regarded as a one-dimensional road network, and the jammed clusters in the highway are therefore two-dimensional (one of space and one of time). Indeed, as can be seen from Fig. \ref{fig:two}\emph{E} and Fig. \ref{fig:two}\emph{F}, the distributions of cluster size and recovery duration of the two-dimensional jammed clusters also show a clear scale free scaling, but with different typical exponents. As seen in {\color{black}{SI Appendix, Fig. S6}}, the scaling exponents are also surprisingly stable and almost do not change from day to day. For the one-dimensional highway, the scaling exponent for traffic resilience is much smaller than that of two-dimensional city, suggesting higher chance of larger jam and longer recover duration. This lower resilience is probably due to the fact that in jammed highways no alternative routing paths are available for traffic flows, while jams in city traffic network have more opportunities to be resolved. On the other hand, as shown in {\color{black}{SI Appendix, Fig. S7}}, resilience of urban traffic in holiday is higher with a higher exponent (2.69 ± 0.06) between Oct. 1st and 7th (the National Day of China) with significantly decreased total traffic demand.

\begin{figure}
\centering
\includegraphics[width=1.0\linewidth]{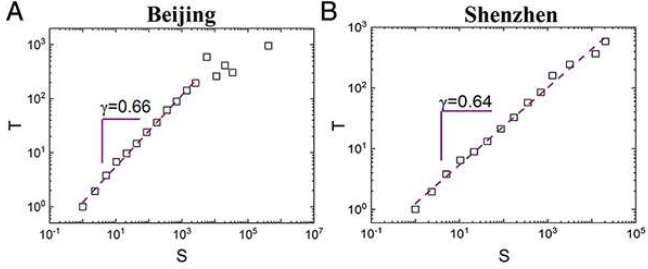}
\caption{\textbf{Recovery time versus cluster size}  (\emph{A}) in Beijing on Oct. 26th 2015 and  (\emph{B}) in Shenzhen on Oct. 26th 2015.}\label{fig:four}
\end{figure}

In order to understand the relationship between traffic jam and recovery duration, we show in Fig. \ref{fig:four}\emph{A} that the recovery time of jammed clusters increases with cluster size with a scaling relation,
\begin{align*}
T \sim S^{\gamma},\numberthis \label{eqn:powerlaw}
\end{align*}
where $\gamma$ is the scaling exponent. This scaling exponent is found similar in both Beijing and Shenzhen. Moreover, this further indicates that the same general mechanism exists for all sizes of jams. For ecological and climate systems, it has been found that the recover rate of system bouncing back from perturbations becomes gradually slow when approaching the tipping point \cite{r44}. While this is rarely observed and confirmed in engineering systems, transportation, as one of the largest complex engineering {\color{black}{systems}}, is observed here the recovery duration grows with the increasing system failure size. We also test the relation between cluster size and recovery duration of jammed clusters of the Beijing-Shenyang Highway (see {\color{black}{SI Appendix, Fig. S8C}}) and find a different power law relation. The value of the $\gamma$ values is also obtained stable for all the observed days as shown in {\color{black}{SI Appendix, Fig. S8.}} {\color{black}{Besides the temporal dimension of the spatio-temporal jammed clusters, we also test the spatial dimension of the resilience clusters (see SI Appendix, Fig. S9) and found that the structures are self-affine and the spatial dimension grows much slower than the temporal dimension.}} 

Next, we ask if these three exponents $\alpha$, $\beta$ and $\gamma$ can be theoretically related. Indeed, if we assume {$P(S) \sim S^{-\alpha }$}, {$\Phi (T) \sim T^{-\beta}$} and {$T \sim S^{\gamma}$} ($\alpha$, $\beta$, $\gamma$ >0), the exponents $\alpha$, $\beta$ and $\gamma$ should be related through the relation between the distributions \cite{r45},
\begin{align*}
P(S)=\Phi (T)\frac{dT}{dS},\numberthis \label{eqn:powerlaw}
\end{align*}
from which we obtain,
\begin{align*}
\gamma=\frac{\alpha -1}{\beta -1}.\numberthis \label{eqn:powerlaw}
\end{align*}
Indeed, Eq. (7) is valid within the error bars found for these exponents (see the comparison of actual value of $\gamma$ with theoretical value, Eq. (7), of $\gamma$ in {\color{black}{SI Appendix, Fig. S8}}). 

\section*{Discussion}

In summary, we have developed a novel and intuitive definition of traffic resilience based on the spatio-temporal evolution of jammed clusters. We find, based on real data that both spatio-temporal cluster size of jams and their recovery duration follow a scale-free distribution, suggesting universal responses of transportation system to different perturbation scenarios. {\color{black}{Note that in the temporal scale, reference \cite{r46} discusses the scaling relation between lifetime of traffic jam and system size in a 1D lattice of cellular automaton (CA) model. In the spatial scale, reference \cite{r47} has also found spatial correlations in traffic flow fluctuations showing a power law decay. While their findings of scaling in spatial or temporal scale are consistent with our results, we identify here a novel combined spatio-temporal scaling of traffic jams. This may help to design mitigation methods viewing jam in a stereo way. For example, similar delay in different cities with different efficiencies \cite{r48}, may be the result of the similar scaling of spatio-temporal congestions formed by the self-organized criticality mechanism.}} These scaling relations are predictable and independent of fluctuating traffic demand from different days in two different cities. The currently absence of suitable definition of traffic resilience could have been the reason for the shortage of efficient allocation of mitigation resources and policy design for risk management. Our findings suggest that urban traffic in different cities could be classified into a few groups, with each group being characterized by the same scaling functions and the same set of scaling exponents. Each group with its intrinsic response to various perturbations, requires different resilience management. Our result is of great theoretical interest, motivating in analogy to critical phenomena and the universality principle, theoretical studies regarding the intriguing question: “Which traffic management variables are critical for determining the resilience scaling functions, and which are irrelevant?”

Moreover, the indications found here in universality of traffic resilience are also of much practical interest. Specifically, when performing resilience management methods \cite{r49,r50,r51}, one may pick the most tractable traffic jam to study, which will help to predict behaviors for all the other jams in the same universality class, especially the likelihood of extreme event by statistical extrapolations \cite{r52}. The relationship between the cluster size and recovery duration can be applied to predict \cite{r53} the congestion influence and the behavior of a certain jam size, which can help the decision-making in the management of transportation. Meanwhile, further studies including model simulations are needed to test and explain the universal characteristics of our results.

{\color{black}{While many studies focus mainly on traffic control in the macroscopic or microscopic scales with dimensionless objectives including travel time or speed \cite{r54} , here we propose a new resilience indicator in the combined spatio-temporal dimension. While it becomes increasingly difficult (if not impossible) to avoid traffic congestions, in the present study, we wish to understand the development and recover behavior of congestion. Our novel method may help to design traffic control method to slow, diminish and shrink the spatio-temporal jammed clusters, leading to improved system resilience. By plotting (see SI Appendix, Fig. S4) day by day for example, the maximal cross section of the spatio-temporal congestion cluster in Beijing, this stable cross section will first help to locate the high-frequency congestion region in real-time traffic. Existing traffic controls are aiming at signal control or road pricing to achieve the optimal operations. For urban traffic, these methods now focus on the global flow properties including Macroscopic Fundamental Diagrams (MFD) \cite{r55}, and rarely consider the spatio-temporal organization of jam in the controlled region, which is the focus of the present manuscript. Using our findings, design schemes and control methods could help to disintegrate the growth of jammed clusters and balance the spatial organization of traffic flow in a more accurate and controlled manner. Furthermore, the scaling laws we identified can help to predict cluster jams above certain values and balance cost and efficiency.}} Future works should also focus, based on our novel approach, on evaluating the traffic resilience in other cities and other infrastructures when appropriate data becomes available. With the broad range of applications of network resilience, developing innovative interdisciplinary approaches based on big data to identify and understand the origin of the scaling laws \cite{r56,r57,r58} of system resilience is thus a future big challenge. 

{\color{black}{A key gradient to achieve the sustainability for a given system is to learn from past failures and enhance the system resilience. In the present study, the resilience definition is mainly applied to traffic congestion scenario, as the major failure of transportation. For other natural and engineering systems, such as ecological damages or communication failures, the relevant resilience can be generalized based on our definition, considering the spatio-temporal features of system adaptation and recover under perturbation. The knowledge of system adaptation and recovery can help to better evaluate the system risk, predict the size of damages in the  system or even collapse and better mitigate against various perturbations.}} \\


\bibliography{Ref}

\end{document}